\documentclass[12pt,preprint]{aastex}

\slugcomment{Accepted by ApJ, October 18 2004}

\begin{document}

\title{Diameters of Mira Stars Measured Simultenously in the J,H,K$'$
Near-Infrared Bands}

\author{ 
R. Millan-Gabet\altaffilmark{1}, E. Pedretti\altaffilmark{2,6},
J. D. Monnier\altaffilmark{2}, F. P. Schloerb\altaffilmark{3},
W. A. Traub\altaffilmark{4}, N. P. Carleton\altaffilmark{4},
M. G. Lacasse\altaffilmark{4} and D. Segransan\altaffilmark{5}}

\email{rafael@ipac.caltech.edu}

\altaffiltext{1}{Michelson Science Center, California Institute of Technology, Pasadena, CA 91125, USA}
\altaffiltext{2}{University of Michigan, Astronomy Department, Ann Arbor, MI 48109, USA}
\altaffiltext{3}{University of Massachusetts at Amherst, Astronomy Department, Amherts, MA 01003, USA}
\altaffiltext{4}{Harvard-Smithsonian Center for Astrophysics, Cambridge, MA 02138, USA}
\altaffiltext{5}{Observatoire de Gen\`{e}ve, Switzerland}
\altaffiltext{6}{NASA Michelson postdoctoral fellow}

\begin{abstract}

We present the first spatially resolved observations of a sample of 23
Mira stars simultaneously measured in the near-infrared J, H and K$'$
bands.  The technique used was optical long baseline interferometry,
and we present for each star visibility amplitude measurements as a
function of wavelength.  We also present characteristic sizes at each
spectral band, obtained by fitting the measured visibilities to a
simple uniform disk model.  This approach reveals the general relation
J~diameter $<$ H~diameter $<$ K$'$~diameter.

\end{abstract}


\section{Introduction}

Mira stars are variable, pulsating stars in the asymptotic giant
branch (AGB). They have pulsation periods of about $100-1000$~days,
during which their light curves exhibit large amplitude variations (as
high as 10~magnitudes at visual wavelengths, becoming a more modest
$\sim~1$~mag in the near-infrared), and their spectral types vary by a
few sub-classes. During the pulsation cycle, their photospheric radii
are thought to vary by $5-50$~\%.  Mira star atmospheres are
geometrically extended and have very low effective temperatures
($\lesssim 3000$ K), hence they are a rich molecular composition. They
exhibit high mass loss rates (a few $10^{-6} M_{\sun}/yr$), and the
ejected material forms layers which become sites of circumstellar
maser activity and dust formation. Thus, although the Mira
evolutionary stage is relatively short ($\sim 5\times10^5$~yrs), it is
a crucial one in order to understand a variety of phenomena such as
high mass loss, the formation of planetary nebulae, and the chemical
enrichment of the galaxy.

Despite the existing large body of observational and theoretical work,
many aspects of the physics of Mira stars remain poorly understood; in
particular their detailed atmospheric stratification and composition,
and their dominant mode of pulsation. This is to a large extent a
consequence of their intrinsic complexity (treatment of 3D convection,
photospheric shocks, and rich photospheric chemistry), and most needed
are new observational data that can help discriminate between the
large number of theoretical models that have been proposed.

Spatially resolved observations of Mira stars are clearly desirable,
and can be easily obtained using high resolution techniques such as
long baseline interferometry. The geometrically extended and
chemically rich nature of Mira star atmospheres manifests itself in
apparent physical sizes which depend on the wavelength of observation,
especially when crossing regions in the spectrum where the opacity
changes rapidly. This fact was first observed in the visible region of
the spectrum, where sizes as much as 50\% smaller are found in
narrowband near-continuum filters than in filters contaminated by TiO
absorption
\citep{bonneau73,labeyrie77,bonneau82,haniff95,hofmann00,hofmann01}.

Although originally expected to be better probes of the underlying
stellar continuum, measurements at infrared wavelengths also exhibit
non-trivial wavelength dependence.  Synthesizing measurements obtained
at different facilities, it appears that as a general trend, Mira
sizes appear large in the visible
\citep{tuthill94,haniff95,ireland04a}, smaller in the near-infrared
\citep[and references therein]{perrin99,hofmann02,gvb02}, and larger
again at thermal infrared wavelengths
\citep{mennesson02,weiner03a,weiner03b}. The interpretation of this
broad trend is that the outer layers (i.e. those layers above a level
that could be called ``the photosphere'') are relatively obscure in
the visible, relatively transparent in the near-infrared, and again
relatively obscuring in the thermal infrared. The physical basis for
this variation of opacity with wavelength can be modelled once the
underlying emprical variation of diameters with wavelength is
established by observations. The purpose of this paper is to show that
there is such a variation through the near-infrared, as defined by the
J, H and K$'$ bands.

Due to instrumental limitations, {\it simultaneous} measurements at
different wavelengths are scarce in the literature, but are clearly
the optimum way to explore the wavelength-dependent structure, since
Mira stars are suspected of exhibiting important cycle-to-cycle
non-repeatibility. Dramatic size increases ($25-100\%$) have been
observed between near ($\sim 1.1-2.3 \mu m$) and thermal ($\sim
3.1-3.8 \mu m$) infrared measurements, both in narrow
\citep{tuthill00a,tuthill00b} and broad bands
\citep{mennesson02}. Smaller but significant ($1-20$\%) effects have
been observed in narrow bands within the near-infrared K-band ($\sim
2.0 - 2.4 \mu m$), as the narrow filters preferentially sample
portions of the spectrum corresponding to strong molecular bands or
the adjacent continuum \citep{thompson02a,thompson02b,perrin04}.
Using the same portion of the spectrum as in this work,
\citet{weigelt03} studied the visibility of the Mira star T~Cep
dispersed across the J/H/K bands, finding similar J and H sizes which
are $\sim 10$\% smaller than the K-band size; as well as size
differences within the K-band as large as 26\%.

Clearly, the richness of the size-wavelength relation holds the
potential to be the best probes of atmospheric structure in these
stars. In turn, as has been illustrated in the recent multi-wavelength
modelling effort by \citet{perrin04}, a complete model that includes
the stellar photosphere, dominant opacity sources and possible
additional emitting layers must be globally fit in order to also
extract accurate stellar diameters, and derive conclusions about the
pulsation mode.

In this paper we present the first systematic survey of Mira star
diameters measured near-simultaneously in the near-infrared J, H and
K$'$ bands. This new multi-color data set allows us in particular to
derive apparent size ratios, which are immune to many sources of
calibration error. In $\S$~\ref{observations} we describe the
measurement technique and Mira sample observed; in $\S$~\ref{analysis}
we describe the data analysis procedure; in $\S$~\ref{results_sec} we
present our results, both in the form of uniform diameter fits at each
spectral band in order to illustrate the main trends in these
wavelength regions, and as tables of calibrated visibility data for
straightforward use in global fitting including other datasets; in
$\S$~\ref{discussion} we discuss our main results and places them in
context of recent theoretical work.

\section{Observations} \label{observations}

Observations were carried out at the Infrared-Optical Telescope Array
(IOTA), a long baseline interferometer located on Mount Hopkins, AZ
\citep[see e.g.][]{Traub98}.  The data presented here were obtained
using the two telescopes available at the time, on two configurations
which provide physical baseline lengths of approximately 21~m and
38~m, at geographic azimuth angles (measured East of North) of
$-4~\deg$ and $+18~\deg$ respectively.

The beam combination and fringe detection instrumentation is exactly
as described in \citet{RMG99a} and \citet{RMG01}.  In short, afocal
telescope beams were combined at a bulk optics beam combiner, and
interferograms were produced by modulating the optical path difference
(OPD) between the two beams by $\pm 60 \mu m$. The fringe modulation
frequencies are in the range $100 - 2000$~Hz, chosen as an optimum
trade-off between maximizing the photon signal and minimizing the
readout time to freeze out atmospherically induced OPD variations
during the fringe acquisition.  The two outputs of the beam splitter
were focused on two well-separated pixels of a NICMOS3 array, the
readout of which is synchronized with the optical path modulation
\citep{RMG99b}.

The NICMOS3 camera contains a set of three standard near-infrared J
($\lambda_0 = 1.25 \mu m, \Delta \lambda = 0.28 \mu m$), H ($\lambda_0
= 1.65 \mu m, \Delta \lambda = 0.30 \mu m$) and K$'$ ($\lambda_0 =
2.16 \mu m, \Delta \lambda = 0.32 \mu m$) filters, where ($\lambda_0$,
$\Delta \lambda$) are the center wavelength and full-width at
half-maximum respectively. Rapid switching between the three filters
guaranteed near-simultaneous observations (within a few minutes) in
the three bands.

A typical observation consisted of 500 interferograms, obtained in a
few minutes, followed by a measurement of the sky background. Target
observations are interleaved with an identical sequence obtained on
calibrator stars which are unresolved by the interferometer or have
small diameters which can be estimated. This sequence was repeated for
each of the three filters.

The basic properties of the Mira stars observed are listed in
Table~\ref{stars}, and a log of observations is given in
Table~\ref{log}.

\section{Data Analysis and Validation} \label{analysis}

The data reduction and calibration procedures are also essentially as
described in \citet{RMG99a} and \citet{RMG01}.  For each
interferogram, a peak fringe visibility (the fringe amplitude relative
to the mean flux detected) is estimated by (a) correcting the
intrinsic detector non-linearity, (b) sky subtraction, (c) subtracting
the normalized outputs of the beam splitter, to enhance the SNR, and
(d) fitting the central three fringes of the resulting interferograms
in the time domain to a point source response template computed as the
Fourier transform of the spectral bandpass used in the observation.

As a result of the fringe fitting procedure, we obtain the following
quantities for each interferogram: mean, fringe amplitude, position of
the central fringe, and sampling rate. The data are flagged as being a
false fringe identification if an individual fringe visibility is
outside the range $0.0-1.0$, or if either the central fringe position
or sampling rate has a value which is more than three times the
standard deviation obtained from the ensemble of the 500 reduced
interferograms in the observation. Finally, an entire observation is
rejected if the standard deviation of the central fringe position
values exceeds $L/2$, where $L=120\,\mu$m is the OPD sweep for each
interferogram.

For each observation, the above procedure results in several hundred
estimates of the fringe visibility, one per interferogram that
survives the flagging procedure. A final estimate for the observation
is formed as the mean of those values, with an internal statistical
error given by the error in the mean.

Calibrator stars are measured close in space and time to the target
observations in order to estimate the instrument's system
visibility. Calibrated visibility amplitudes are obtained by dividing
the target visibilities by estimates of the system visibility at the
time of the target observations formed by spline interpolation of the
calibrator observations. A correction to the system visibility
resulting from the finite but known size of the calibrator stars is
applied. Standard error propagation is used to include in the
calibrated visibility the error resulting from the division by the
system visibility and the systematic error resulting from the
uncertainty in the calibrator star sizes.  

The calibrator sizes and uncertainties are in Table~\ref{cals}; these
values were estimated as follows: (1) if available, adopt a
near-infrared direct measurement from the CHARM catalog~\citep{charm};
or else (2) if available, adopt the value based on stellar atmosphere
modelling of \citet{borde02} or \citet{cohen99}; or else (3) if
available, and if the star is a luminosity class I, II, or III, use
the empirical size-color relations of \citet{gvb99}; or else (4) fit
the available visual and infrared photometry with a stellar blackbody
model, using the {\it Fbol} module of the {\it getCal}
software~\footnote{{\it getCal} is a a software package for planning
of interferometer observations produced by the Michelson Science
Center at the California Institute of Technology ({\tt
http://msc.caltech.edu/software/index.html)}}.

In order to obtain characteristic sizes at each wavelength, the
calibrated visibility data obtained at each spectral band were fit
using a simple uniform disk (UD) model for the stellar brightness.
Although it is well known that for Mira stars the UD is in most cases
not a good representation of the true center-to-limb variation (CLV),
reconstruction of the true CLV is not possible when only sparsely
sampled visibility data are available. In near-continuum bandpasses
however, as near-infrared filters are to a large extent, UD profiles
are expected to approximate the CLV better than at other wavelengths
\citep{scholz03}. Moreover, all the visibility data presented here
were obtained at spatial frequencies well inside the first lobe of the
visibility function (except for one star, R~Leo, as discussed in
$\S$~\ref{results_sec}), where departures from the UD intensity
profile are virtually undistinguishable, and therefore baseline
dependent errors are expected to be minimal in this dataset.

\begin{figure}
\begin{center}
\includegraphics[angle=90,scale=0.7]{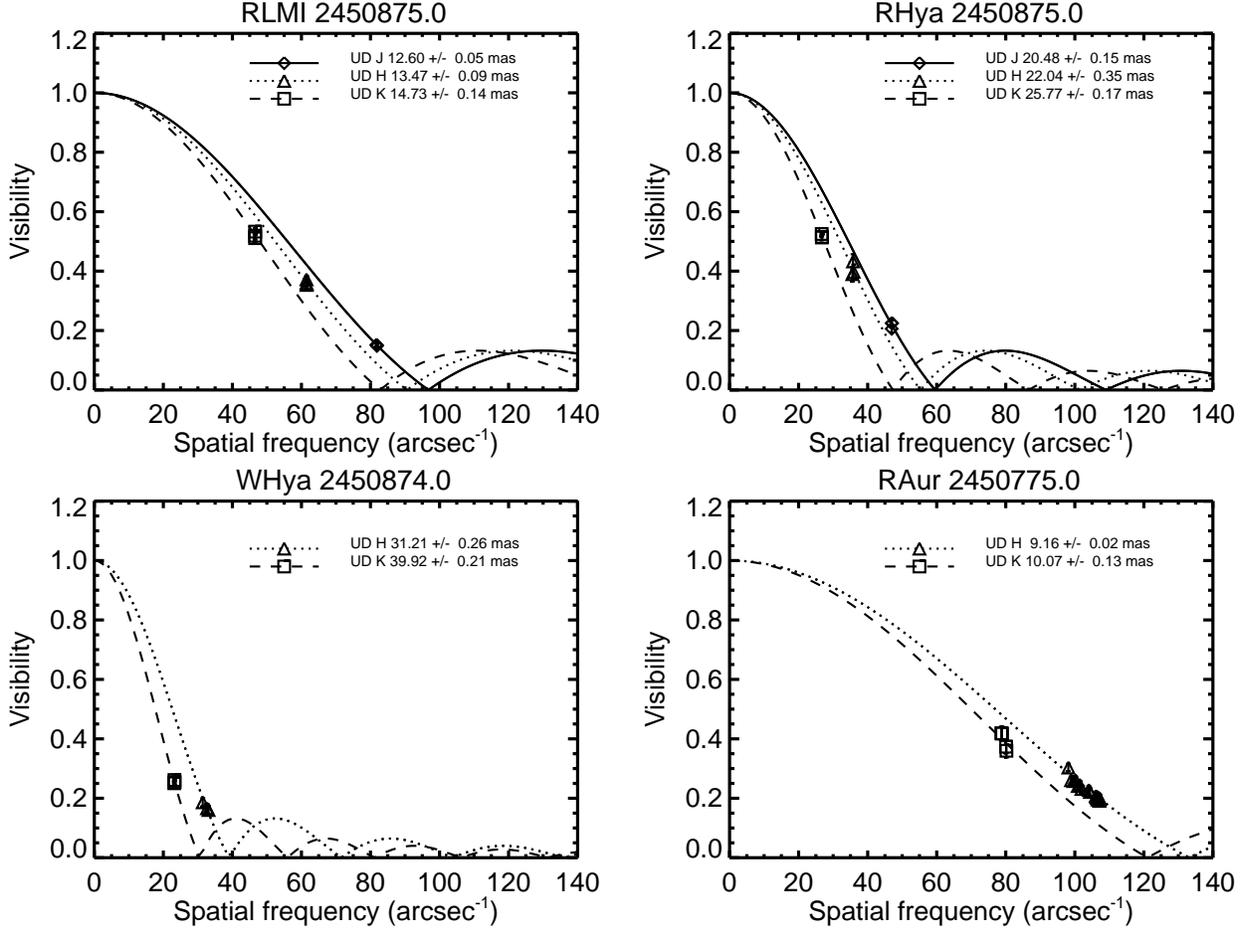}
\caption{
Example visibility data. The four panels illustrate the general
finding that lower visibilities are measured at the longer
near-infrared wavelengths (J~diameter $<$ H~diameter $<$
K$'$~diameter). The bottom-right panel also illustrates the data
precision. In all panels, the abscissa is the spatial frequency given
by the projected baseline $B_p/\lambda_0 = \sqrt{u^2+v^2}$; and the
ordinate is the visibility amplitude (1.0 for a point-like source,
0.0 for a completely resolved extended source).\label{examples}}
\end{center}
\end{figure}

In Figure~\ref{examples}, we show 4 representative examples (out of a
total of 34 similar observations) of the data and fitting results. The
examples are chosen to illustrate distinct aspects of the data. The
top two panels illustrate the general trend found that J~diameter $<$
H~diameter $<$ K$'$~diameter. The bottom left panel illustrates our
ability to sample relatively low visibilities for bright sources.  The
bottom right panel shows an example where Earth rotation allowed
sampling of a relatively large range of projected baseline length,
illustrating the visibility data precision for observations within a
single night; here the visibility fit RMS is 0.01, or 3.8\% of the
(changing) model visibility, which is a typical value for the dataset
(full range is 1$-$10\%).

The night-to-night data consistency was evaluated by comparing the
results obtained when the same star was observed on different nights,
at similar pulsational phase and using the same baseline, in order to
isolate the test from possible size variations and non-circular star
shapes. There are 3 such cases in our dataset ($\times$ 3 bands for a
total of 9 independent comparisons) and the result is that the
night-to-night consistency in visibility is in all cases better than
8\%, with mean values being 2.3\%, 3.0\% and 3.8\% for J, H and K$'$
respectively.

Our data can also be compared with previous measurements from the
literature. This comparison can only be made at K-band. Considering
the most recent measurements \citep{perrin99,gvb02,monnier04} we find
that (a) our diameters show no systematic positive or negative bias
with respect to the literature values, and (b) the diameter values
agree to 20\% or better. Considering that no effort was made to
compare data taken only at similar pulsational phases, this level of
agreement appears reasonable, given for example that K-band Mira sizes
can change by 10\% for a phase difference of only 0.04
\citep{perrin99}.

\section{Results} \label{results_sec}

Recognizing that potentially the best use of our J/H/K$'$ measurements
will be as part of a larger modelling effort which incorporates for
each star all the available data across the spectrum, a goal beyond
the scope of this paper, we provide in Table~\ref{vis_data} the
totality of our visibility measurements. Columns~$1-4$ give the star
name, calendar/Julian observation epoch, and filters
used. Columns~$5-7$ give the 2-dimensional spatial frequencies
($u,v$), and measured visibilities.

In addition, in order to capture the essential features of the
``size''$-$wavelength relation across the near-infrared, we provide in
Table~\ref{results} the results of fitting our visibility data to
UD models.  Columns~$1-4$ give the star name, observation Julian date,
the pulsation phase~\footnote{For each observation we have estimated
the pulsation phase at the observation epoch as follows. The AFOEV
({\tt http://cdsweb.u-strasbg.fr/afoev/}) and AAVSO ({\tt
http://www.aavso.org/}) databases were used to obtain the epochs of
maximum light before and after the IOTA observation, $o_{max1}$ and
$o_{max2}$ respectively. The phase at the IOTA epoch is then estimated
as $\mbox{Phase} = (\mbox{IOTA epoch} - o_{max1}) /(o_{max2} -
o_{max1})$.} and an estimate of the spectral type for that pulsation
phase (see $\S$~\ref{discussion}). Columns~$5-7$ give the J, H and
K$'$-band best fit UD diameters. The fact that we have measurements at
3 different wavelengths, allows us to make a relative measurement in
terms of diameter {\it ratios}, given in columns~$8-9$, which are
expected to be immune to some sources of systematic error
(e.g. calibrator diameter uncertainty, seeing effects) and also
represent the linear size ratios independent of (uncertain) distances.

For R~Leo, essentially a single spatial frequency was sampled at H and
K$'$ bands, and the visibilities are sufficiently low that
mathematically, three solutions to the UD fit exist: one in the main
lobe of the visibility function, and two in the second lobe. We
therefore use apriori knowledge of R~Leo's nominal near-infrared size
to further constrain our solutions.  For K$'$-band, we use the
diameter measured by \citet{monnier04} ($30.3 \pm 0.2$~mas) as a
starting value for the fit. For H-band, we use the unpublished value
obtained by one of us (J. D. Monnier) from aperture masking at the
Keck telescope ($\sim 30$~mas), to obtain the two first lobe solutions
indicated in Table~\ref{results}. Without spatial frequency coverage
that follows the shape of the visibility function, it is not possible
to discriminate between these solutions, and therefore we have droped
this star from the analysis that follows.

As is apparent in Table~\ref{results}, the basic result of this paper
is that the near-infrared diameters are found to follow the trend:
J~diameter $<$ H~diameter $<$ K$'$~diameter. This result is also
illustrated in Figure~\ref{ratios}, where the angular diameter ratios
are plotted as a function of pulsation phase.  The mean values of the
diameter ratios are:

\begin{eqnarray}
\overline{R_{J/H}} & = & 0.93 \pm 0.02 \nonumber \\
\overline{R_{H/K'}} & = & 0.89 \pm 0.02 \nonumber \\
\end{eqnarray}

\noindent where the $1-\sigma$ uncertainty is given by the error in
the mean, and the (sample RMS, number of observations) are (0.09,16)
and (0.09,24) for J/H and H/K$'$ respectively.

We finally comment on a few cases for which there are intriguing
changes in the measured sizes, above the levels of calibration
accuracy demonstrated in $\S$~\ref{analysis}.  As can be seen in
Table~\ref{results}, whenever there is both a significant time
difference between two epochs of observation and the baselines used
were different, apparent size changes in the range $10-30$\% can occur
(R~Aur, U~Ori, R~Lmi, S~Crb, R~Ser).  Given the variety of possible
effects at play however (physical changes in the stars, inadequacy of
the UD model, stellar asymmetries probed for different baseline
orientations), it is difficult to assess whether these effects are
real or arise from unusually high calibration errors. These issues
might be resolved with further observations.

\section{Discussion} \label{discussion}

We have measured the diameters of 23 Mira stars simultaneously in the
near-continuum near-infrared J, H and K$'$ bands, a first for each
star in the sample. The simultaneity of the measurements guarantees
that the same physical state of the star is measured in each band. We
find the apparent size relation: J~diameter $<$ H~diameter $<$
K$'$~diameter, in agreement with expectations that the shorter
near-infrared filter (J-band) better samples the true photospheric
continuum, and that the H and K$'$ filters contain progressively more
opacity contamination from molecules in higher layers \citep[see
e.g.][]{jacob02}. In terms of the magnitude of the size difference, or
diameter ratios, we find that the stars appear to be as much as 24\%
smaller at J than at H (mean 8\%) and as much as 27\% smaller at H
than at K$'$ (mean 11\%). These magnitudes are in general agreement
with the models of \citet{jacob02} and \citet{ireland04b}; although
clearly detailed comparisons must be made for specific stars and
specific models.

\begin{figure}
\begin{center}
\includegraphics[angle=90,scale=0.7]{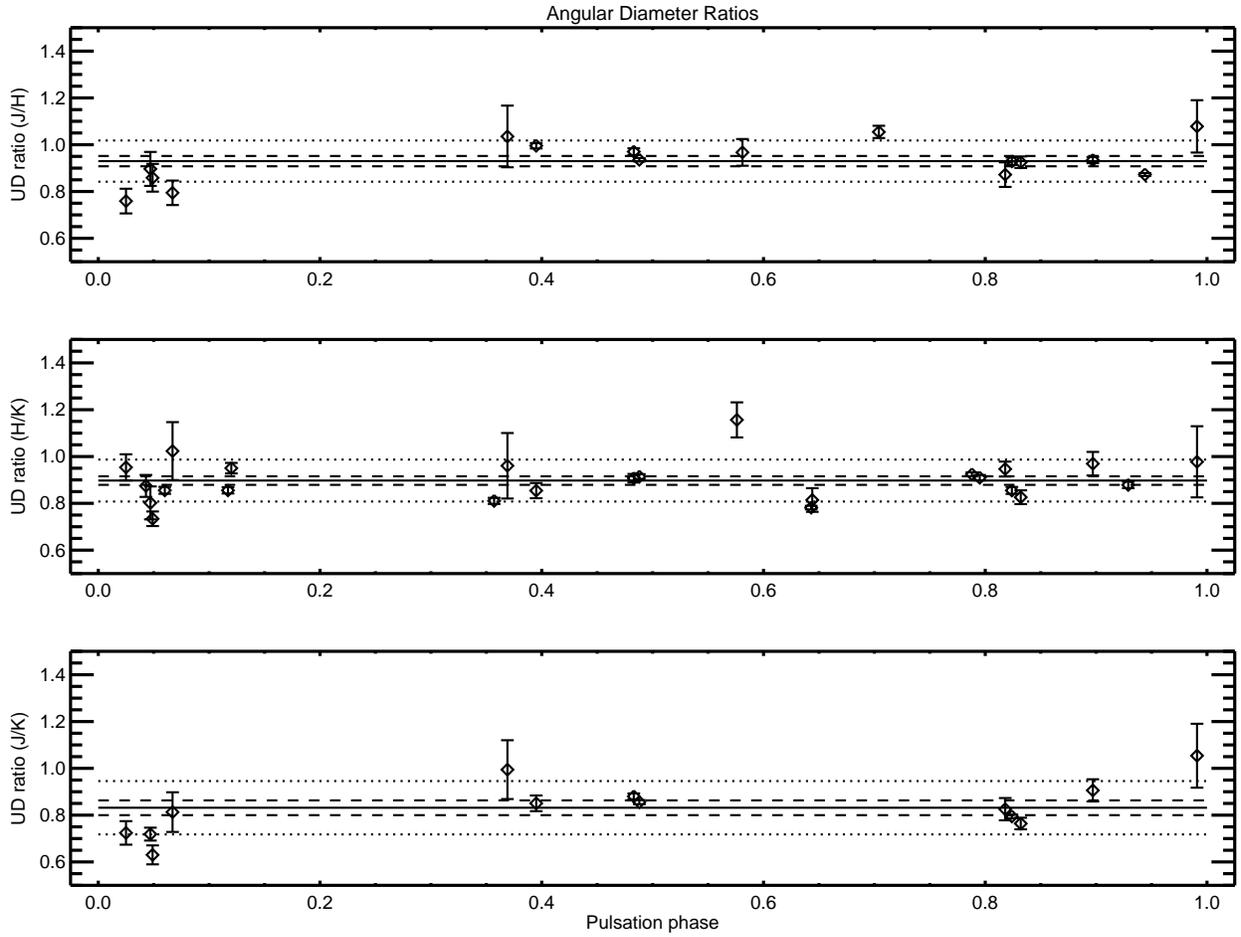}
\caption{
Angular diameter ratios. For each panel, the solid line indicates the
measured mean ratio, the dashed lines indicate the error in the mean,
and the dotted lines indicate the RMS of the sample.\label{ratios}}
\end{center}
\end{figure}

Changes in the near-infrared sizes of individual Mira stars as a
function of pulsation phase have been detected interferometrically
\citep{perrin99,thompson02a,thompson02b}. No single star in our sample
has sufficient phase coverage for such a study. However, we have
attempted to detect the pulsation by treating the sample as a single
``synthetic'' Mira. A priori, the most promising approach appeared to
be to explore this effect using the size ratios, due to their
calibration advantages. However, as is apparent in
Figure~\ref{ratios}, no sinusoidal signature is present.  We have also
explored the possibility that the pulsation was not seen in the above
approach due to the fact that instantaneous spectral type, rather than
pulsation phase, might be the relevant parameter. Thus, we have used
``phased spectral types'' (column~4 of Table~\ref{results}) from the
literature \citep{terrill69,lockwood71,lockwood72} to search for a
monotonic relation (also for the ``synthetic Mira'') between the
measured apparent sizes and the spectral type measured for the
corresponding pulsation phase, also without success.

\begin{figure}
\begin{center}
\includegraphics[angle=90,scale=0.7]{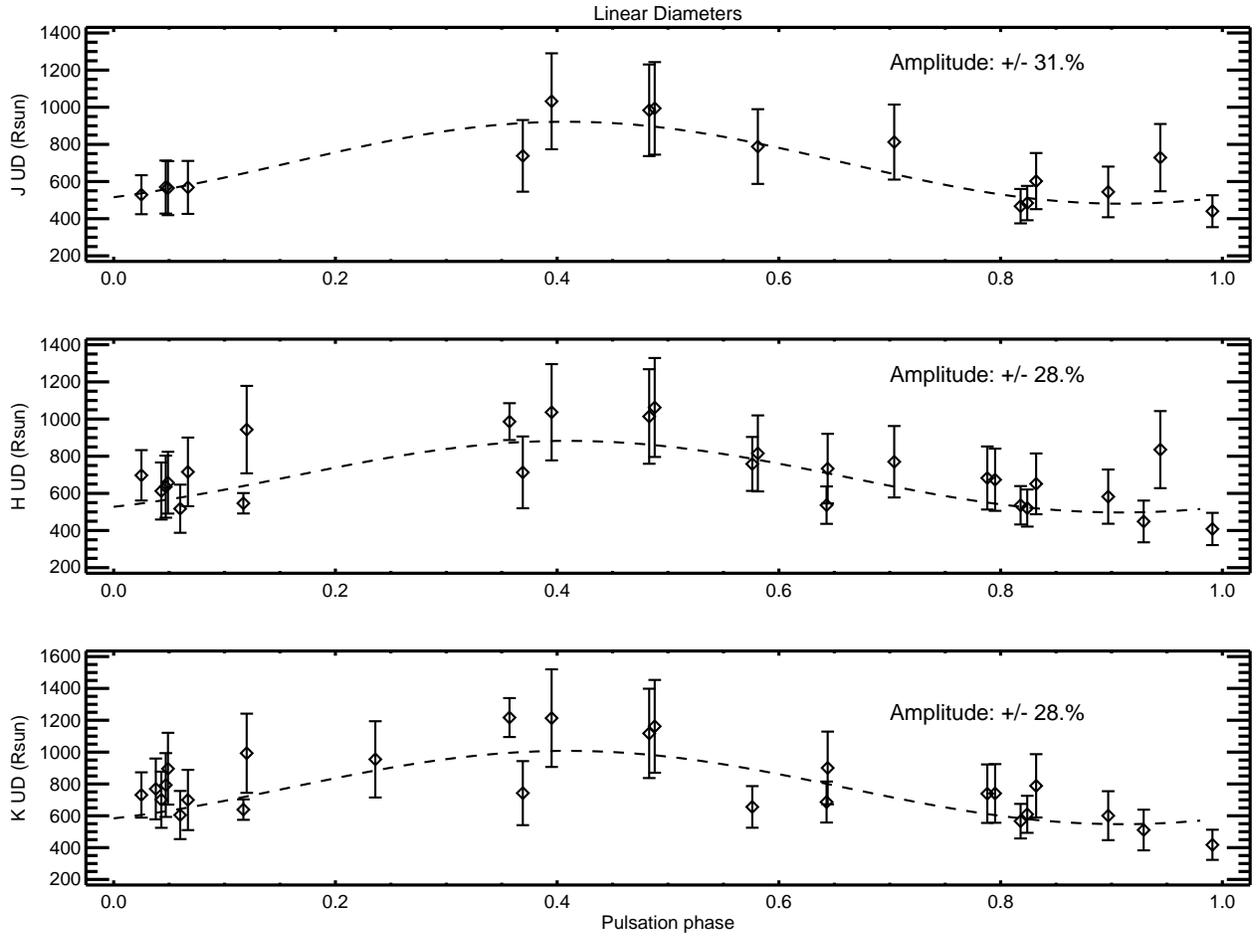}
\caption{
Linear diameters as a function of pulsation phase. In each panel, the
dashed line indicates a sinusoidal fit to the data. The sinusoidal
fits have 4 free parameters: mean, amplitude, period and phase; except
for the K$'$-band data, where we fix the period and phase to the
values found in the J and H fits.  See $\S$~\ref{discussion} for
discussion.
\label{diams}}
\end{center}
\end{figure}

Forsaking the advantages of distance independence then, we show in
Figure~\ref{diams} our measured J/H/K$'$ linear sizes as a function of
pulsation phase.  As dicussed by \citet{gvb02}, Hipparcos distances to
Mira stars are rather uncertain, and therefore we have used distances
from the literature based on period-luminosity (PL) relations, as
indicated in Table~\ref{stars}.  We note that we do not attempt to
apply the PL relations using estimates of the absolute K-band
magnitudes based on our own fringe data because the instrumentation
and observing strategy were never intended for accurate
photometry. Although the data errors are relatively large, and
systematic deviations are apparent, the data suggests a sinusoidal
signature with maximum sizes near phase $=$ 0.5 (minimum light), as
expected (as the photospheric size increases and cools, the star
becomes fainter).  The statistical significance of the sinusoidal
signature is however marginal, with reduced $\chi^2$ of 0.4, 0.8 and
0.9 at J/H/K$'$ respectively, compared with 1.9, 2.0 and 2.2 for
models consisting of the constant average diameters. We note that
while the observed sinusoid amplitudes ($50 - 60$\% peak-to-peak) are
comparable to those obtained from model predictions
\citep{ireland04b}, they are considerably larger than observed ($9 -
26$\% peak-to-peak) when individual Miras are followed through their
pulsation cycles \citep{thompson02a,thompson02b}. Also, it can be seen
in Figure~\ref{diams}, that the linear sizes vary rather in unison,
explaining why the pulsation was not apparent in the diameter ratios
of Figure~\ref{ratios}.

Ultimately, we believe that the failure to more convincingly detect
the pulsation using the approaches just described is due to the
fundamental inadequacy of those methods, as several physical
mechanisms may contribute dispersion at levels comparable to the
expected signatures: (a) intrinsic star-to-star differences, (b) our
filters are broadband, and therefore probe a complex combination of
emission from multiple layers of molecular gas just above the
photosphere, with opacities which also change during the pulsation,
making the true expected signature very complicated, and in particular
potentially masking the photospheric layer motions \citep{ireland04b},
and (c) some theoretical models predict cycle-to-cycle
non-repeatibility of the photospheric diameters, at significant levels
\citep[tens of \%;][]{jacob02,ireland04b}, which if present would
introduce that much dispersion in our apparent size vs. phase
relations. On that last point, we note however that for the only
published multi-cycle measurements to date
\citep{thompson02a,thompson02b} this effect was not observed; although
it appears to be present (at $5-15$\% levels) in yet un-published data
by the same group (Thompson R. R. 2004, private communication).  In
the literature, the success of the ``synthetic'' Mira method is mixed:
while \citet{gvb96} saw a 20\% peak-to-peak change in the effective
temperatures of a sample of 18 Miras, the effect was not explored in
the 22 star sample of their follow-up paper \citet{gvb02}, and appears
not to be present by straighforward plotting of their published
numbers.  We finally note that the best case in our sample in which to
search for a pulsation signature of an individual star (U~Per, period
$=$ 320 days) was observed at two epochs almost exactly one cycle
apart, and the measured UD sizes differ by 10, 12, and 6\% at J, H and
K$'$ respectively, a marginally significant size change.

Several optical interferometers around the world are now equipped to
measure accurate visibilities in broad or narrow spectral bands, and
together span the visible to mid-infrared spectrum. Although recent
efforts over more limited regions of the spectrum have clearly
illustrated the power of multi-wavelength spatially resolved
observations \citep{perrin04,jacob04,ohnaka04,weiner04}, global
modelling fitting has however not been attempted to date, but
represents the necessary next step toward a better theoretical
understanding of Mira star atmospheres and dynamics.

\acknowledgments

The authors wish to acknowledge fruitful discussions with Robert
Thompson, Gerard van Belle and Sam Ragland. This work has made use of
services produced by the Michelson Science Center at the California
Institute of Technology. This research has made use of the SIMBAD
database, operated at CDS, Strasbourg, France.  This research has made
use of the NASA/ IPAC Infrared Science Archive, which is operated by
the Jet Propulsion Laboratory, California Institute of Technology,
under contract with the National Aeronautics and Space
Administration. This research has made use of NASA's Astrophysics Data
System Service. E.P. wishes to acknowledge that part of this work was
performed while he was a predoctoral fellow of the Smithsonian
Astrophysical Observatory.

\clearpage

\begin{deluxetable}{lcccrrrccc}
\rotate
\tablecolumns{10}
\tablewidth{0pc}
\tablecaption{Basic properties of Mira star sample. \label{stars}}
\tablehead{
\colhead{Name} & \colhead{RA} & \colhead{DEC} & \colhead{V$_{min}-$V$_{max}$} & 
\colhead{J} & \colhead{H} & \colhead{K} & \colhead{Spectral} & \colhead{Distance} & \colhead{Distance} \\
 & (J2000) & (J2000) & & & & & Type & (pc) & reference 
}
\startdata
R And  & 00 24 01.95 & $+$38 34 37.3 & 7 $-$ 15 & 2.0     & 0.7     & 0.1     & S  & 532 $\pm$ 101 & 2 \\
Z Cet  & 01 06 45.13 & $-$01 28 52.9 & 9 $-$ 14 & 4.6     & 3.7     & 3.3     & M5 & 920 $\pm$ 175 & 4 \\ 
U Per  & 01 59 35.12 & $+$54 49 20.0 & 8 $-$ 12 & 2.2     & 1.2     & 0.9     & M6 & 559 $\pm$ 140 & 1 \\
R Per  & 03 30 02.98 & $+$35 40 17.1 & 9 $-$ 14 & 4.8     & 4.0     & 3.2     & M4 & \nodata       & \nodata \\
R Aur  & 05 17 17.69 & $+$53 35 10.0 & 7 $-$ 13 & 1.0     & -0.1    & $-$0.5  & M7 & 342 $\pm$ 85  & 1 \\
S Ori  & 05 29 00.89 & $-$04 41 32.7 & 8 $-$ 14 & 0.9     & $-$0.01 & $-$0.5  & M7 & 481 $\pm$ 120 & 1 \\
U Ori  & 05 55 49.17 & $+$20 10 30.7 & 6 $-$ 13 & 1.1     & 0.2     & $-$0.3  & M8 & 310 $\pm$ 77  & 1 \\
X Aur  & 06 12 13.38 & $+$50 13 40.4 & 8 $-$ 14 & 4.5     & 3.8     & 3.2     & K2 & \nodata       & \nodata \\
R Cnc  & 08 16 33.83 & $+$11 43 34.5 & 6 $-$ 11 & 0.8     & $-$0.3  & $-$0.7  & M7 & 300 $\pm$ 30  & 5 \\
R LMi  & 09 45 34.28 & $+$34 30 42.8 & 8 $-$ 13 & 1.7     & 0.5     & $-$0.1  & M7 & 367 $\pm$ 22  & 1 \\
R Leo  & 09 47 33.49 & $+$11 25 43.6 & 6 $-$ 10 & $-$0.7  & $-$1.7  & $-$2.3  & M8 & 115 $\pm$ 29  & 1 \\ 
R Hya  & 13 29 42.78 & $-$23 16 52.8 & 5 $-$ \phn 8  & $-$1.3  & $-$2.2  & $-$2.7  & M7 & 110 $\pm$ 21  & 4 \\ 
W Hya  & 13 49 02.00 & $-$28 22 03.5 & 6 $-$ \phn 9  & $-$1.7  & $-$2.6  & $-$3.2  & M7 & 80  $\pm$ 15  & 4 \\ 
S Crb  & 15 21 23.96 & $+$31 22 02.6 & 7 $-$ 13 & 1.1     & 0.2     & $-$0.2  & M7 & 375 $\pm$ 94  & 1 \\
RS Lib & 15 24 19.79 & $-$22 54 39.9 & 7 $-$ 13 & 1.0     & $-$0.1  & $-$0.6  & M  & 210 $\pm$ 40  & 3 \\
R Ser  & 15 50 41.73 & $+$15 08 01.1 & 6 $-$ 14 & 1.3     & 0.4     & 0.1     & M7 & 356 $\pm$ 89  & 1 \\ 
S Her  & 16 51 53.92 & $+$14 56 30.8 & 6 $-$ 13 & 2.3     & 1.2     & 1.0     & M6 & 677 $\pm$ 169 & 1 \\
Ry Lyr & 18 44 51.9  & $+$34 40 30.0 & 10 $-$ 15 & \nodata & \nodata & \nodata & M6 & \nodata       & \nodata \\
R Aql  & 19 06 22.2  & $+$08 13 48.0 & 6 $-$ 12 & 0.7     & $-$0.3  & $-$0.8  & M7 & 224 $\pm$ 56  & 1 \\
R Peg  & 23 06 39.2  & $+$10 32 36.1 & 7 $-$ 13 & 1.8     & 0.9     & 0.4     & M7 & 435 $\pm$ 109 & 1 \\
W Peg  & 23 19 50.5  & $+$26 16 43.7 & 8 $-$ 13 & 1.1     & 0.2     & $-$0.1  & M7 & 310 $\pm$ 31  & 5 \\
S Peg  & 23 20 32.6  & $+$08 55 08.1 & 8 $-$ 13 & 2.3     & 1.4     & 1.0     & M6 & 675 $\pm$ 169 & 1 \\
R Aqr  & 23 43 49.5  & $-$15 17 04.2 & 6.5 $-$ 11 & $-$0.1  & $-$1.1  & $-$1.6  & M7 & 272 $\pm$ 68  & 1 \\
\enddata
\tablecomments{V magnitudes and spectral types are from the AAVSO and SIMBAD databases, 
near-infrared magnitudes are from the 2MASS database; as these stars
are by definition variable, this information is meant to be merely representative.}
\tablerefs{
(1) \citet{gvb02}; 
(2) \citet{gvb97};
(3) \citet{young95};
(4)  \citet{jura92};
(5)  \citet{wyatt83}}

\end{deluxetable}


\clearpage
\begin{deluxetable}{lccccc}
\tablecolumns{6}
\tablewidth{0pc}
\tablecaption{Log of observations. \label{log}}
\tablehead{
\colhead{Name} & \colhead{Date} & \colhead{Julian Date} & \colhead{Baseline} & 
\colhead{Filters} & \colhead{Calibrator(s)} \\
 & \colhead{(yy/mm/dd)} & \colhead{($-$2450000)} & & & }
\startdata
R And	& 97/11/21 & 0774.0 & S15N35 & H/K$'$	& HR82, HR175 \\	
Z Cet 	& 98/11/04 & 1122.0 & S15N35 & J/H/K$'$	& HR353       \\	
U Per 	& 97/11/19 & 0772.0 & S15N35 & J/H/K$'$	& HR787	      \\	
U Per 	& 98/10/01 & 1088.0 & S15N35 & J/H/K$'$	& HR470       \\	
R Per 	& 98/10/02 & 1089.0 & S15N35 & J/H/K$'$	& HR876       \\
R Per 	& 98/11/05 & 1123.0 & S15N35 & J/H/K$'$	& HR876       \\
R Aur 	& 97/10/12 & 0734.0 & S15N15 & J/H	& HR1866      \\
R Aur 	& 97/11/19 & 0772.0 & S15N35 & H/K$'$	& HR1588, HR1866 \\
R Aur 	& 97/11/22 & 0775.0 & S15N35 & H/K$'$	& HR1588 \\
S Ori 	& 98/11/05 & 1123.0 & S15N35 & H/K$'$	& HR1830 \\
U Ori 	& 97/10/17 & 0739.0 & S15N15 & J/H	& HR2169 \\
U Ori 	& 97/11/21 & 0774.0 & S15N35 & K$'$	& HR2047 \\
U Ori 	& 97/11/23 & 0776.0 & S15N35 & H/K$'$	& HR2047 \\
X Aur 	& 97/10/14 & 0736.0 & S15N15 & J/H/K$'$	& HR2338 \\
R Cnc 	& 98/03/02 & 0875.0 & S15N15 & H/K$'$	& HR2864 \\
R Lmi 	& 97/11/20 & 0773.0 & S15N35 & K$'$	& HR3791 \\
R Lmi 	& 98/02/28 & 0873.0 & S15N15 & J/H/K$'$	& HR4081 \\
R Lmi 	& 98/03/02 & 0875.0 & S15N15 & J/H/K$'$	& HR4081 \\
R Leo 	& 98/03/01 & 0874.0 & S15N15 & H/K$'$	& HR3877 \\
R Hya	& 98/02/28 & 0873.0 & S15N15 & J/H/K$'$	& HR5020, HR4958 \\ 
R Hya 	& 98/03/02 & 0875.0 & S15N15 & J/H/K$'$	& HR5020, HR4958 \\
W Hya 	& 98/03/01 & 0874.0 & S15N15 & H/K$'$	& HR5312 \\
S Crb 	& 98/03/01 & 0874.0 & S15N15 & J/H/K$'$	& HR5674 \\
S Crb 	& 98/06/12 & 0977.0 & S15N35 & H/K$'$	& HR5674 \\
RS Lib 	& 98/02/28 & 0873.0 & S15N15 & J/H/K$'$	& HR5824 \\
R Ser 	& 98/03/02 & 0875.0 & S15N15 & J/H/K$'$	& HR5940 \\
R Ser 	& 98/03/07 & 0880.0 & S15N35 & H/K$'$	& HR5940 \\
S Her 	& 98/06/15 & 0980.0 & S15N35 & J/H	& HR6542, HR6065 \\
Ry Lyr 	& 98/06/15 & 0980.0 & S15N35 & H/K$'$	& HR7192 \\
R Aql 	& 98/06/13 & 0978.0 & S15N35 & H/K$'$	& HR7557 \\
R Peg 	& 98/10/01 & 1088.0 & S15N35 & J/H/K$'$	& HR8608 \\
W Peg 	& 98/10/01 & 1088.0 & S15N35 & H/K$'$	& HR9051 \\
S Peg 	& 98/09/30 & 1087.0 & S15N35 & J/H/K$'$	& HR8916 \\
R Aqr  	& 97/10/17 & 0739.0 & S15N15 & J/H/K$'$	& HR8980 \\
\enddata
\tablecomments{The baseline designations are as follows: SxNy refers
to South telescope being at the x~meter station and the North telescope
at the y~meter station. The [length,azimuth] of the S15N15 and S15N35
baselines are [21~m,$-$3.8~deg] and [38~m,$+$18~deg] respectively.}

\end{deluxetable}
\clearpage

\begin{deluxetable}{llccc}
\tablecolumns{5}
\tablewidth{0pc}
\tablecaption{Calibrator estimated diameters and uncertainties. \label{cals}}
\tablehead{
\colhead{Calibrator} & \colhead{Calibrator} & \colhead{Spectral Type} & 
\colhead{Angular Diameter} & \colhead{Reference} \\ 
(HR number) & (HD number) & & (mas) & }
\startdata
HR82	& HD1671   & F5III  & 0.65 $\pm$ 0.08 &   1a  \\  
HR175   & HD3817   & G8III  & 1.19 $\pm$ 0.15 &	 3   \\
HR353   & HD7147   & K4III  & 1.59 $\pm$ 0.20 &	 3   \\
HR470   & HD10110  & K5III  & 1.84 $\pm$ 0.02 &	 2   \\
HR787   & HD16735  & K0II   & 1.29 $\pm$ 0.16 &	 3   \\
HR876   & HD18339  & K3III  & 1.45 $\pm$ 0.18 &	 3   \\
HR978   & HD20277  & G8IV   & 1.14 $\pm$ 0.09 &	 4   \\
HR1133  & HD23193  & A2m    & 0.23 $\pm$ 0.04 &	 4   \\
HR1588  & HD31579  & K4III  & 1.78 $\pm$ 0.23 &	 3   \\
HR1698  & HD33856  & K3III  & 2.13 $\pm$ 0.02 &	 1b  \\
HR1830  & HD36134  & K1III  & 1.29 $\pm$ 0.16 &	 3   \\
HR1866  & HD36678  & M0III  & 2.55 $\pm$ 0.03 &	 2   \\
HR2047  & HD39587  & G0V    & 1.11 $\pm$ 0.20 &	 4   \\
HR2105  & HD40486  & K0     & 1.87 $\pm$ 1.33 &	 4   \\
HR2169  & HD42049  & K4III  & 2.17 $\pm$ 0.28 &	 3   \\
HR2188  & HD42466  & K1III  & 0.96 $\pm$ 0.12 &	 3   \\
HR2338  & HD45466  & K4III  & 2.17 $\pm$ 0.28 &	 3   \\
HR2864  & HD59294  & K1III  & 2.24 $\pm$ 0.03 &	 2   \\
HR3115  & HD65522  & K2     & 1.50 $\pm$ 0.98 &	 4   \\
HR3319  & HD71250  & M3III  & 4.61 $\pm$ 0.60 &	 3   \\
HR3791  & HD82522  & K4III  & 1.13 $\pm$ 0.14 &	 3   \\
HR3877  & HD84561  & K4III  & 2.15 $\pm$ 0.27 &	 3   \\
HR4081  & HD90040  & K1III  & 1.48 $\pm$ 0.19 &	 3   \\
HR4958  & HD114149 & K0III  & 1.51 $\pm$ 0.19 &	 3   \\
HR5020  & HD115659 & G8III  & 3.27 $\pm$ 0.41 &	 3   \\
HR5228  & HD121156 & K2III  & 0.96 $\pm$ 0.12 &	 3   \\
HR5312  & HD124206 & K3III  & 1.78 $\pm$ 0.22 &	 3   \\
HR5674  & HD135438 & K5     & 2.57 $\pm$ 2.70 &	 4   \\
HR5824  & HD139663 & K3III  & 1.99 $\pm$ 0.02 &	 2   \\
HR5940  & HD142980 & K1IV   & 1.27 $\pm$ 0.16 &	 3   \\
HR6065  & HD146388 & K3III  & 1.22 $\pm$ 0.15 &	 3   \\
HR6542  & HD159353 & K0III  & 1.10 $\pm$ 0.14 &	 3   \\
HR6770  & HD165760 & G8III  & 1.74 $\pm$ 0.22 &	 3   \\
HR6793  & HD166229 & K2III  & 1.56 $\pm$ 0.20 &	 3   \\
HR7192  & HD176670 & K3III  & 2.33 $\pm$ 0.03 &	 2   \\
HR7557  & HD187642 & A7V    & 3.25 $\pm$ 0.41 &	 1c  \\
HR7804  & HD194258 & M5III  & 6.62 $\pm$ 0.87 &	 3   \\
HR8608  & HD214298 & K5     & 1.77 $\pm$ 0.87 &	 4   \\
HR8916  & HD220954 & K1III  & 1.94 $\pm$ 0.02 &	 2   \\
HR8961  & HD222107 & G8III  & 2.64 $\pm$ 0.33 &	 3   \\
HR8980  & HD222547 & K4III  & 2.73 $\pm$ 0.35 &	 3   \\
HR8987  & HD222643 & K3III  & 2.00 $\pm$ 0.25 &	 3   \\
HR9010  & HD223173 & K3II   & 2.48 $\pm$ 0.32 &	 3   \\
HR9029  & HD223559 & K4III  & 2.15 $\pm$ 0.27 &	 3   \\
HR9051  & HD224128 & K5     & 1.49 $\pm$ 0.55 &	 4   \\
\enddata
\tablerefs{
(1) from CHARM catalog \citep{charm};
(1a) \citet{lane01}; 
(1b) \citet{cohen99};
(1c) \citet{gvb01};
(2)  \citet{borde02};
(3)  \citet{gvb99};
(4)  {\it getCal/Fbol}}

\end{deluxetable}


\clearpage
%
\begin{deluxetable}{lccccccc}
\tablecolumns{8}
\tablewidth{0pc}
\tablecaption{Calibrated visibility data. \label{vis_data}}
\tablehead{
\colhead{Name} & \colhead{Date} & \colhead{Julian Date} & \colhead{Filter} & \colhead{$u$} & \colhead{$v$} & 
\colhead{Visibility}\\
 & \colhead{(yymmdd)} & & & \colhead{($arcsec^{-1}$)} & \colhead{($arcsec^{-1}$)} & Amplitude}
\startdata
R And   &       971121    &     2450774     &     H     &      -2.22    &      112.25     &      0.463     $\pm$      0.035  \\
R And   &       971121    &     2450774     &     H     &      -4.79    &      112.16     &      0.419     $\pm$      0.035  \\
R And   &       971121    &     2450774     &     H     &     -12.91    &      111.48     &      0.485     $\pm$      0.035  \\
R And   &       971121    &     2450774     &     K$'$  &     -13.46    &       83.05     &      0.766     $\pm$      0.040  \\
R And   &       971121    &     2450774     &     K$'$  &     -15.92    &       82.57     &      0.712     $\pm$      0.040  \\
Z Cet   &       981104    &     2451122     &     J     &      57.79    &      118.31     &      0.858     $\pm$      0.026  \\
Z Cet   &       981104    &     2451122     &     J     &      54.73    &      118.25     &      0.850     $\pm$      0.026  \\
Z Cet   &       981104    &     2451122     &     J     &      51.98    &      118.19     &      0.863     $\pm$      0.026  \\
Z Cet   &       981104    &     2451122     &     H     &      52.04    &       89.86     &      0.858     $\pm$      0.021  \\
Z Cet   &       981104    &     2451122     &     H     &      50.23    &       89.80     &      0.843     $\pm$      0.021  \\
Z Cet   &       981104    &     2451122     &     H     &      48.24    &       89.74     &      0.838     $\pm$      0.021  \\
Z Cet   &       981104    &     2451122     &     K$'$  &      26.08    &       67.09     &      0.913     $\pm$      0.010  \\
Z Cet   &       981104    &     2451122     &     K$'$  &      24.98    &       67.07     &      0.918     $\pm$      0.010  \\
U Per   &       971119    &     2450772     &     J     &     -10.34    &      140.05     &      0.509     $\pm$      0.022  \\
U Per   &       971119    &     2450772     &     J     &     -14.74    &      139.53     &      0.515     $\pm$      0.022  \\
\enddata
\tablecomments{
The complete version of this table is in the electronic edition of the
Journal.  The printed edition contains only a sample.  The complete
data set is also available in the OI-FITS format \citep{pauls04} upon
request.}

\end{deluxetable}

\clearpage
%
%
\begin{deluxetable}{lclclllll}
\rotate
\tablecolumns{9}
\tablewidth{0pc}
\tablecaption{Near-infrared angular diameters and diameter ratios. \label{results}}
\tablehead{
\colhead{Name} & \colhead{Julian Date} & \colhead{Phase} & \colhead{Phased} & 
\colhead{$D_J$} & \colhead{$D_H$} &  \colhead{$D_K$}  & 
\colhead{$R_{J/H}$} & \colhead{$R_{H/K}$} \\
 & \colhead{(-245000)} & & \colhead{Spectral Type} & \colhead{(mas)} & \colhead{(mas)} & \colhead{(mas)} & & \\
 & & & \colhead{(M subclass)} & & & & & }
\startdata
   R And & 0774.0 & 0.6  & 6.0 &   \nodata &  6.6 $\pm$  0.2 &  5.7 $\pm$  0.3 &   \nodata &  1.16 $\pm$  0.07  \\
   Z Cet & 1122.0 & 0.25 & 5.0 &  2.7 $\pm$  0.2 &  3.5 $\pm$  0.2 &  3.7 $\pm$  0.2 &  0.76 $\pm$  0.05 &  0.95 $\pm$  0.06  \\
   U Per & 0772.0 & 0.8  & 6.0 &  5.0 $\pm$  0.1 &  5.4 $\pm$  0.1 &  6.6 $\pm$  0.2 &  0.93 $\pm$  0.02 &  0.83 $\pm$  0.03  \\
   U Per & 1088.0 & 0.9  & 5.5 &  4.5 $\pm$  0.1 &  4.85 $\pm$  0.05 &  5.0 $\pm$  0.3 &  0.93 $\pm$  0.01 &  0.97 $\pm$  0.05  \\
   R Per & 1089.0 & 0.3  & 6.3 & $<$  1.6 & $<$  0.9 & $<$  2.7 & \nodata & \nodata  \\ 
   R Per & 1123.0 & 0.5  & 7.5 & $<$  1.5 & $<$  2.2 & $<$  1.8 & \nodata & \nodata  \\ 
   R Aur & 0734.0 & 0.7  & 8.5 & 11.1 $\pm$  0.1 & 10.5 $\pm$  0.3 &   \nodata &  1.05 $\pm$  0.03 &   \nodata \\
   R Aur & 0772.0 & 0.8  & 8.5 &   \nodata &  9.3 $\pm$  0.1 & 10.1 $\pm$  0.1 &   \nodata &  0.92 $\pm$  0.01  \\
   R Aur & 0775.0 & 0.8  & 8.5 &   \nodata &  9.2 $\pm$  0.1 & 10.1 $\pm$  0.1 &   \nodata &  0.91 $\pm$  0.01 \\
   S Ori & 1123.0 & 0.1  & 6.3 &   \nodata &  9.1 $\pm$  0.1 &  9.6 $\pm$  0.2 &   \nodata &  0.95 $\pm$  0.02 \\
   U Ori & 0739.0 & 0.9  & 9.0 & 10.9 $\pm$  0.1 & 12.5 $\pm$  0.1 &   \nodata &  0.87 $\pm$  0.01 &   \nodata  \\
   U Ori & 0774.0 & 0.04 & 8.2 & 11.5 $\pm$  0.1 & \nodata & \nodata & \nodata & \nodata  \\ 
   U Ori & 0776.0 & 0.04 & 8.2 &   \nodata &  9.2 $\pm$  0.3 & 10.5 $\pm$  0.4 &   \nodata &  0.87 $\pm$  0.05  \\ 
   X Aur & 0736.0 & 0.1  & 3.4 & $<$  3.3 & $<$  3.2 &  $<$  5.2 & \nodata & \nodata  \\ 
   R Cnc & 0875.0 & 0.4  & 7.5 &   \nodata & 15.3 $\pm$  0.2 & 18.9 $\pm$  0.2 &   \nodata &  0.81 $\pm$  0.01  \\
   R LMi & 0773.0 & 0.2  & 8.0 &   \nodata &   \nodata & 12.1 $\pm$  0.1 &   \nodata &   \nodata  \\
   R LMi & 0873.0 & 0.5  & 9.0 & 12.5 $\pm$  0.1 & 12.8 $\pm$  0.2 & 14.2 $\pm$  0.2 &  0.97 $\pm$  0.01 &  0.91 $\pm$  0.02  \\
   R LMi & 0875.0 & 0.5  & 9.0 & 12.6 $\pm$  0.1 & 13.5 $\pm$  0.1 & 14.7 $\pm$  0.1 &  0.94 $\pm$  0.01 &  0.91 $\pm$  0.01  \\
   R Leo & 0874.0 & 0.4  & 7.5 &   \nodata & 23.8 $\pm$  0.3  & 29.91 $\pm$  0.27 & \nodata &  0.79 $\pm$  0.01  \\	
   \nodata & \nodata &\nodata & \nodata & \nodata & 32.4 $\pm$  0.4  & \nodata & \nodata & \nodata                    \\                
   R Hya & 0873.0 & 0.8  & 8.0 & 19.8 $\pm$  1.1 & 22.7 $\pm$  0.6 & 23.9 $\pm$  0.5 &  0.87 $\pm$  0.05 &  0.95 $\pm$  0.03  \\
   R Hya & 0875.0 & 0.8  & 8.0 & 20.5 $\pm$  0.2 & 22.0 $\pm$  0.4 & 25.8 $\pm$  0.2 &  0.93 $\pm$  0.02 &  0.86 $\pm$  0.01  \\
   W Hya & 0874.0 & 0.6  & 7.0 &   \nodata & 31.2 $\pm$  0.3 & 39.9 $\pm$  0.2 &   \nodata &  0.78 $\pm$  0.01  \\
   S Crb & 0874.0 & 0.4  & 7.5 &  9.2 $\pm$  0.7 &  8.9 $\pm$  0.9 &  9.2 $\pm$  0.9 &  1.04 $\pm$  0.13 &  0.96 $\pm$  0.14  \\
   S Crb & 0977.0 & 0.6  & 9.0 &   \nodata &  9.1 $\pm$  0.5 & 11.2 $\pm$  0.4 &   \nodata &  0.81 $\pm$  0.05  \\
  RS Lib & 0873.0 & 1.0  & 7.5 &  9.8 $\pm$  0.4 &  9.1 $\pm$  0.9 &  9.3 $\pm$  1.1 &  1.08 $\pm$  0.11 &  0.98 $\pm$  0.15  \\
   R Ser & 0875.0 & 0.0  & 4.0 &  7.5 $\pm$  0.1 &  8.3 $\pm$  0.7 & 10.4 $\pm$  0.4 &  0.90 $\pm$  0.07 &  0.80 $\pm$  0.07  \\
   R Ser & 0880.0 & 0.1  & 6.0 &   \nodata &  6.8 $\pm$  0.1 &  7.9 $\pm$  0.1 &   \nodata &  0.86 $\pm$  0.01  \\
   S Her & 0980.0 & 0.6  & 8.0 &  5.4 $\pm$  0.3 &  5.6 $\pm$  0.2 &   \nodata &  0.97 $\pm$  0.06 &   \nodata  \\
  Ry Lyr & 0980.0 & 0.35 & 7.7 & \nodata &     2.7 $\pm$ 0.3 & $<$  2.03 & \nodata & \nodata  \\ 
   R Aql & 0978.0 & 0.9  & 7.5 &   \nodata &  9.3 $\pm$  0.1 & 10.6 $\pm$  0.1 &   \nodata &  0.88 $\pm$  0.01  \\
   R Peg & 1088.0 & 0.0  & 7.0 &  6.0 $\pm$  0.4 &  7.0 $\pm$  0.3 &  9.6 $\pm$  0.2 &  0.86 $\pm$  0.06 &  0.73 $\pm$  0.03  \\
   W Peg & 1088.0 & 0.1  & 6.9 &   \nodata &  8.2 $\pm$  0.1 &  9.6 $\pm$  0.1 &   \nodata &  0.86 $\pm$  0.01  \\
   S Peg & 1087.0 & 0.1  & 5.2 &  3.9 $\pm$  0.1 &  4.9 $\pm$  0.3 &  4.8 $\pm$  0.5 &  0.79 $\pm$  0.05 &  1.02 $\pm$  0.12  \\
   R Aqr & 0739.0 & 0.4  & 7.5 & 17.7 $\pm$  0.2 & 17.7 $\pm$  0.1 & 20.8 $\pm$  0.8 &  1.00 $\pm$  0.01 &  0.85 $\pm$  0.03  \\
\enddata
\tablecomments{For R~Leo, the two H-band diameter values indicated correspond to the two possible solutions that exist
in the second lobe of the visibility function, see $\S$~\ref{results_sec} for discussion.}

\end{deluxetable}



\clearpage

\clearpage

\clearpage

\end{document}